\documentclass[apjl]{emulateapj}

\shorttitle{On the modulation of RR Lyrae stars in the globular cluster M3}
\shortauthors{Jurcsik et al.}

\begin{document}

\title{On the modulation of RR Lyrae stars in the globular cluster M3}

\author{J. Jurcsik\altaffilmark{1}, P. Smitola\altaffilmark{1}, G. Hajdu\altaffilmark{2,3} and J. Nuspl\altaffilmark{1}}

\affil{$^1$ Konkoly Observatory, H-1525 Budapest PO Box 67, Hungary}
\affil{$^2$ Instituto de Astrof\'{i}sica, Pontificia Universidad Cat\'olica de Chile,
Av. Vicu\~na Mackenna 4860, 782-0436 Macul, Santiago, Chile}
\affil{$^3$ Millennium Institute of Astrophysics, Av. Vicu\~na Mackenna 4860,
782-0436 Macul, Santiago, Chile}

\email{jurcsik@konkoly.hu}

\begin{abstract}
A new, extended time-series photometry of M3 RR Lyrae stars has revealed that four of the ten double-mode stars show large-amplitude Blazhko modulation of both radial modes. The first, detailed analysis of the peculiar behavior of the unique, Blazhko RRd stars is given. While the P1/P0 period ratio is normal, and the overtone mode is dominant in the other RRd stars of the cluster,  the period ratio is anomalous and the fundamental mode has a larger (or similar) mean amplitude than the overtone has in Blazhko RRd stars. The modulations of the fundamental and overtone modes are synchronized only in one of the Blazhko RRd stars. No evidence of any connection between the modulations of the modes in the other three stars is found. The Blazhko modulation accounts, at least partly, for the previously reported amplitude and period changes of these stars.
Contrary to the $\sim50$\% Blazhko statistics of  RRab and RRd stars, Blazhko modulation occurs only in 10\% of the overtone variables in M3. Four of the five Blazhko RRc stars are bright, evolved objects, and one has a similar period and brightness as Blazhko RRd stars have.  The regions of the instability strip with high and low occurrence rate of the Blazhko modulation overlap with the regions populated by first- and second-generation stars according to theoretical and observational studies, raising up the possibility that the Blazhko modulation occurs preferentially in first-generation RR Lyrae stars.

\end{abstract}


\section{Introduction}

M3 is one of the globular clusters (GC) rich in double-mode RR Lyrae (RRd) stars, some of them showing a peculiar behavior. The puzzling modal changes of V79, which had been an RRab star before 1992, was a double-mode star with a dominant overtone mode between 1992 and 2007, and has been a Blazhko RRab star since 2007,  was analyzed and discussed in details by  \cite{cl97}, \cite{cl99}, \cite{v79ct} and \cite{go10}. In a comprehensive study, \cite{cc04} did not find a fully convincing explanation for the anomalously small period ratios of V13 and V200. They concluded:``that both mass dispersion and strong evolutionary effects seem to be present" and that ``it may well be that the pulsation models and the theoretical Petersen diagram do not adequately predict masses for variable stars undergoing rapid evolutionary processes".  M3 is also unique in that the fundamental mode is dominant in  three  RRd stars, while the dominant mode is the overtone in the wast majority of the known RRd stars.

The Blazhko effect \citep{bl}, the amplitude and/or phase modulation observed in about 50\% of the RRab stars \citep{kbs,kepler} is still one of the unsolved problems of the pulsation of RR Lyrae stars.  Among many other ideas which try to explain the phenomenon, time to time, the interaction of near-resonant  radial modes is also raised  \citep{bor,mos,kg,g13,bry}. The second overtone has indeed been detected in many Blazhko RRab stars \citep{m13,kepler} and the first overtone was proposed to explain signals at 0.731, 0.753 and 0.721 frequency ratios in the Kepler data of V445 Lyr, RR Lyr and V360 Lyr, respectively \citep{445,mks12,kepler}, though the identification of this frequency  in V360 Lyr is ambiguous.

The light curves of most of the RRd stars can be completely described with two radial-mode frequencies, their harmonics and linear-combination terms. Minor additional components \citep{aql} or rapid  period and amplitude changes  \citep[mode-switch,][]{cl97,c94,so14a,pol,so14b} in RRd stars were also detected. 

The OGLE-II data showed that a significant fraction of the 1O/2O double-mode Cepheids in the LMC show amplitude and phase modulations of the light curves \citep{mk09}. 
Both radial modes of these stars were modulated with the same modulation period, and the amplitude variation of the modes were anti-correlated. 
The phenomenon was  explained as the non-stationary resonant coupling of one of the radial modes with another radial or non-radial mode,  and with the sharing on the excitation sources between the two radial modes. Blazhko-like behavior of double mode RR Lyrae stars has been announced  based on the OGLE IV survey of the Galactic bulge just very recently \citep{so14b}. 
 

In this letter, we analyze four double-mode stars showing the Blazhko effect in details, based on a new photometry of M3. The distribution and statistics of Blazhko RR Lyrae stars in M3 are also discussed.

\section{Data}

An extended photometric campaign to observe M3  was conducted at the Konkoly Observatory in 2012 using the 90/60 Schmidt telescope (Piszk\'estet\H o). Data will be published in Jurcsik et al (in preparation). Accurate light curves of 160, 42 and ten RRab, RRc and RRd stars were obtained, respectively. The flux curves \citep{isis} of about one-third of the stars are not magnitude calibrated as the zero point of the flux-magnitude transformation is uncertain because of crowding.

\section{Double-mode stars showing Blazhko modulation in M3}

\begin{figure}
\includegraphics[width=9.2 cm]{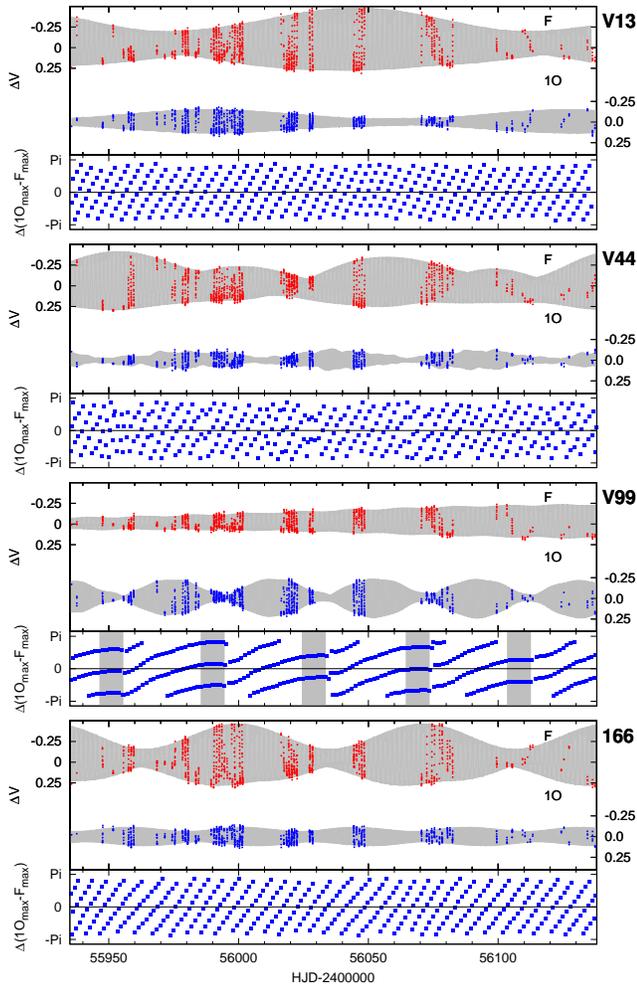}
\caption{Fundamental- and overtone-mode light curves of Blazhko double-mode stars in M3. Using the complete Fourier solutions of the observed light curves, the data are prewhitened for the other mode, its modulation, and for the linear-combination terms. Synthetic data of the modulated light curves of the modes are drawn in gray. The modulation periods of the F and 1O modes are the same only in V13. The bottom panels plot the phase differences of the maxima of the overtone mode and the nearest maxima of the fundamental mode for each star. The shaded regions in the phase-difference plot of V99  highlights the intervals when P1/P0 is the closest to 3/4.}
\label{df}
\end{figure}
\begin{figure}
\includegraphics[width=9.2 cm]{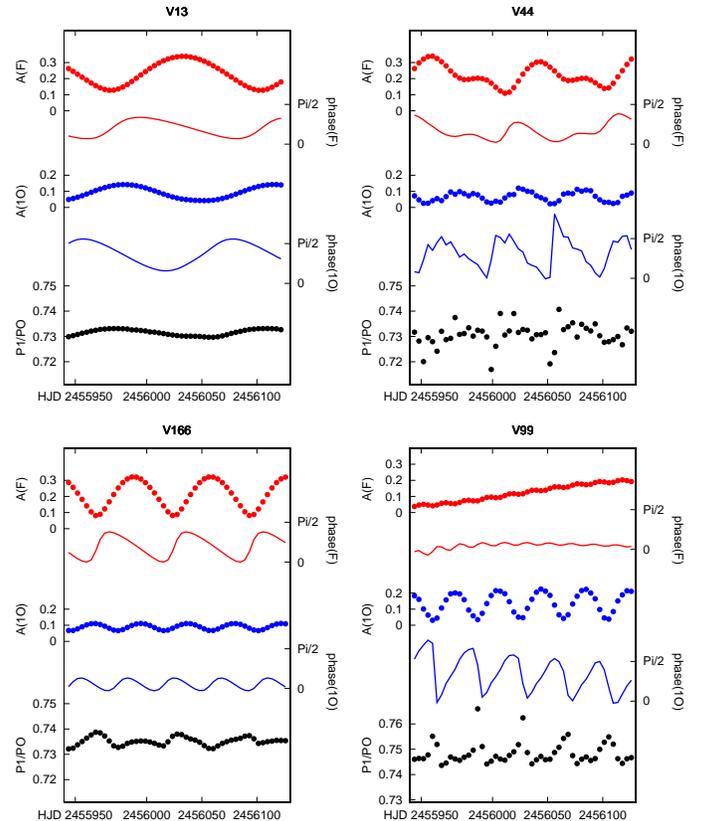}
\caption{The amplitude and the phase variations of the modes and the temporal P0/P1 period ratio derived from the synthetic data are shown for the four Blazhko RRd stars. The modulations of the F and 1O modes of V13 are anticorrelated, no connection between the modulation properties of the modes in the other stars is evident.  }
\label{ered}
\end{figure}

Nine double-mode stars have been identified previously in M3, but one of them, V79 is  a Blazhko RRab star currently. Based on the recent data, two new RRd stars have been detected. 
The overtone mode also appears in the spectrum of V44, a Blazhko RRab star  at a longer fundamental-mode period than what the other RRd stars have, showing complex modulation properties. Most probably, the large amplitude and the complexity of the modulation, as well as the lack of a dense, extended photometric data-set hindered the detection of the fundamental mode in the previous data.  The other new RRd star is V125, a normal RRc star showing a small amplitude signal at the fundamental-mode frequency, as well. 

Besides the pulsation components and linear-combination terms, large-amplitude signals around the radial-mode frequencies were detected in three known double-mode stars (V13, V99 and V166). Thus, together with V44, four  Blazhko RRd stars are identified. The Fourier spectra indicate that both radial modes are modulated on a regular or on a complex way in these stars.  Table~\ref{bl.tab} lists the pulsation and the main modulation periods of the Blazhko RRd stars. The light curves of their fundamental- and overtone-modes  are shown in Figure~\ref{df}. The data are prewhitened for the pulsation and the modulation frequencies of the other mode and for all the linear-combination terms. The synthetic data of the modulated light curves are drawn in gray.

The amplitude and phase variations of the modes and the temporal P1/P0 period ratio derived from the synthetic data are shown in Figure~\ref{ered}.  Both the amplitude- and the phase-modulation of the  modes are anti-correlated in V13.  This type of behavior is the same as observed in the Blazhko 1O/2O Cepheids \citep{mk09}.  The other three stars show, however, quite diverse modulation properties of the  radial modes. We find no evidence of an influence of the radial modes and their modulations on each other. The only common features are that  both modes are modulated in each Blazhko RRd star; the rapid phase-change (either positive or negative) is connected to the amplitude-minimum of the respective mode (a typical feature of Blazhko stars); and that the amplitude and amplitude variation of the fundamental mode are larger (or the same for V99) than those of the overtone mode.  

The phase differences of the  maxima of the overtone mode and the nearest maxima of the fundamental mode  are plotted in the bottom panels of Figure~\ref{df}. As the period ratios  do not equal to 3:4, depending on the value of $P1/P0$, the phase differences of the modes are  similar after each  $dt=P1/(1-1.25*P1/P0)$  cycle. Although the phase variation of the modes perturbs these cycle lengths for some extent, both the phase-difference plots in Figure~\ref{df} and the $dt$ values given in Table~\ref{bl.tab} suggest that the modulation periods are not in connection with the  cycle lengths of the onset of similar, close-resonance conditions. 

The most interesting star is V99, because its P1/P0 period ratio is very close to 0.75. As its overtone mode is strongly phase modulated, the temporal period ratio is even larger than 3:4 around the minimum amplitude phase of the overtone. When this happens (shaded in gray in Figure~\ref{df}) the phase differences of the modes are constant for a while. However, the fixation of the modes occurs at different phase-relations in the different cycles. Even when the modes are locked at the times of the maxima (one of the possible phase differences is around zero), a new Blazhko cycle starts and the phase lock is soon dissolved. Still in this case,  the (near-resonant) radial mode do not seem to influence the pulsation and/or modulation of the other mode.

\begin{table}
\caption{Properties of Blazhko double-mode stars in M3}
\label{bl.tab}
\begin{tabular}{lcrcrll}
\hline
Star&  $P0$  & $P0_{mod}^{a}$ &$P1$  & $P1_{mod}$ & P1/P0 & dt\\
\hline
V13  & 0.47949 & 139(1)  & 0.35072 & 139(1)  & 0.7314& 18.0\\
V44  & 0.50377 &  97.0(2)& 0.36812 &  56.0(8)& 0.7307& 17.3 \\
V99  & 0.48209 & $>450$  & 0.36113 &  40.0(2)& 0.7490& 333 \\
V166 & 0.48504 &  71.5(3)& 0.35672 &  44.0(5)& 0.7355& 23.0\\
\hline
\multicolumn{7}{l}{\footnotesize{$^{a}$ The errors of the last digits are given in parenthesis.}}\\
\end{tabular}
\end{table}

\begin{figure}
\includegraphics[width=8.8 cm]{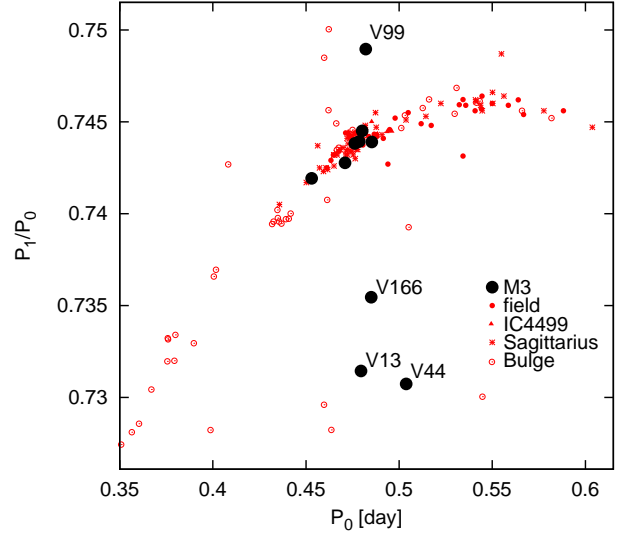}
\caption{The Petersen diagram, $P_1/P_0$ period ratio vs $P_0$, of M3 double-mode stars are shown. For comparison, the period ratios of other known RRd stars are also plotted. }
\label{dm.figure}
\end{figure}

In Figure~\ref{dm.figure}, the $P_1/P_0$ period ratios of M3 RRd stars are shown in comparison with some samples of RRd stars \citep[compilation by][]{mcc,wils,wn,cs,so11}.

The non-Blazhko RRd stars (V68, V87, V125, V200, V251 and V252) have normal period ratio, but the four Blazhko RRd stars have anomalous values. The period ratios are 0.008-0.014 smaller and 0.005 larger than normal for V13, V44, V166 and for V99, respectively. According to pulsation models,  period ratios this low are only possible for stars with high metallicity ($Z > 0.0015$) and/or with strongly reduced mass values \citep{cc04,so11}. To explain the too large period ratio of V99, other sign of anomalies were needed. It is quite unlikely, however, that four of the ten RRd stars in M3 would indeed have anomalous mass/metallicity values.

RRd stars with similarly peculiar period ratios have been detected only in the galactic bulge \citep{so11}. The recent finding that many RRd stars in the bulge also show the Blazhko effect \citep{so14b} suggests that, maybe, Blazhko RRd stars  have  anomalous period ratios in the bulge, as well.
\section{Distribution of Blazhko stars}

\begin{figure}
\includegraphics[width=9 cm]{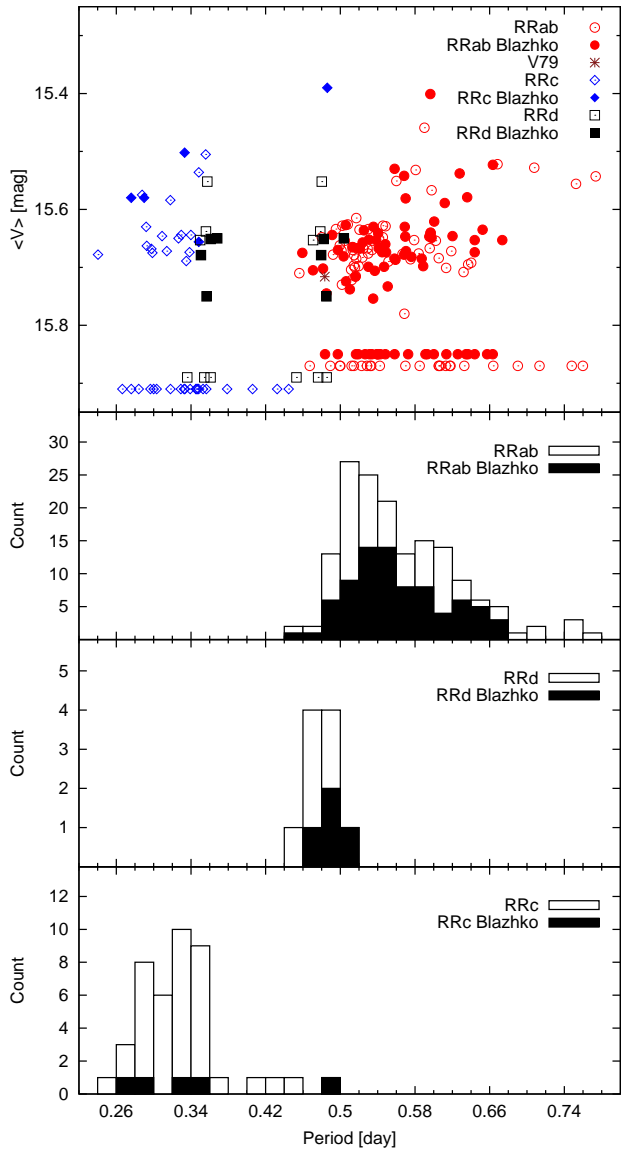}
\caption{{\it Top panel:} the $P$ vs $<V>_{int}$ plot of fundamental-mode (circles), double-mode (squares) and overtone-mode (diamonds)  stars is shown. Blazhko stars are denoted by filled symbols. Double-mode stars are shown twice, both at the fundamental- and at the overtone-mode periods. The periods of stars without magnitude-calibrated light curves are indicated at arbitrary (15.8-15.9) magnitude values.
{\it Three bottom panels:} the histograms of the period distributions of RRab, RRd and RRc stars are shown. The black parts of the histograms denote Blazhko stars. The fundamental-mode periods of RRd stars are used in the statistics. V79 is regarded as a Blazhko RRab star here.}
\label{hist.figure}
\end{figure}

The distribution of Blazhko stars among fundamental, double-mode and overtone variables  in the $P - <V>_{int}$ plane and the histograms of the Blazhko statistics are shown in Figure~\ref{hist.figure}.

The Blazhko statistics of RRab stars are 50\% in M3 (Jurcsik et al in preparation).
Both normal brightness (OoI) and over-luminous (OoII) RRab stars show light-curve modulation. 

Discarding V79, the Blazhko occurrence rate is 4/10 for RRd stars; together with V79, it is 5/11. The mean $V$ magnitudes of the Blazhko RRd stars are close to the brightness of the  Zero-Age Horizontal Branch (ZAHB), and this is true for V79, as well. However, as only one bright RRd star is detected  (V87),  the statistics of  evolved RRd stars are senseless. 
Residual signals at the pulsation frequencies are detected only in five of the 42 observed overtone stars. 
Four of them are over-luminous; the Blazhko statistics are again 50\% (4/8) in the subsample of evolved, bright overtone stars. The fifth overtone Blazhko star (V140) is very close in period and  brightness to the RRd stars. 
Without the evolved variables, only about 3\% (1/34) of the RRc stars show the Blazhko effect.
It seems that the Blazhko instability does not act in overtone variables close to their ZAHB position in M3.
\\

\section{Discussion and conclusion}

There has been a long-time debate whether the modal content of RRd stars in M3 is changing \cite[][]{cl97,cc04,be06}. 
\cite{cc04} concluded that ``... both the extent and actual role of evolution on the period/mode variations of the RRd stars still remain unclear ... if evolution is not the culprit of the mode switching observed in the M3 RRd's, as suggested by \cite{ps}, one must invoke some unknown instability that may be responsible for the changes in period ratios, amplitude ratios, and/or dominant pulsation modes". On the contrary, based on data simulations, \cite{be06} argued that the modal content of RRd stars are stable, and the ``formerly published mode-switching events were the over-interpretation of badly sampled observations". 

The present study has revealed that the light curves of each RRd star with a contradictory historical record of the amplitudes/periods of the modes are Blazhko modulated. Because of the incomplete Blazhko-phase coverage of the previous,  not enough extended and/or sparse photometric data-sets, the amplitude-  and phase-modulations of the modes have led to 
a large spread in the measured amplitudes/periods of the modes. Instead of evolutionary or mode-switching episodes, the Blazhko modulation is the ``unknown instability" that explains most of the observed changes.

\cite{cc04} failed to give a comprehensive explanation for the anomalously low, 0.738 and 0.739 period ratios of V13 and V200. The recent data have made it possible to determine the period ratios with a high accuracy. The results have shown that the period ratio of V200 is normal, but it is as small as 0.731 in V13. The other three Blazhko RRd stars have  anomalous period ratios, too. It is important to note that not only the period ratio is low, but the fundamental-mode period is longer than expected for a normal RRd star in V44 . This is also the case for the three Blazhko RRab stars which probably show the first overtone mode  at anomalous period ratio in the Kepler data \citep{445,mks12,kepler}. 
Extreme values of the fundamental-mode periods and of the period ratios of RRd stars are observed exclusively in stars showing the Blazhko effect. 

V119/M3, a normal RRab star had not shown light-curve modulation in the earlier photometric data, but a strong amplitude modulation was detected in the 2009 \citep{oc} and in the currently concerned 2012 observations. The fundamental-mode period was 0.0002~d shorter according to the 2009-2012 data than it had been previously. The mean amplitude of the modulated light curve is the same as the amplitude of the stable-light-curve data. Therefore, a nonlinear shift does not explain the period decrease of V119.  A similar order of  pulsation-period increase (0.0003 d) has been accompanied the cease of the modulation of OGLE-BLG-RRLYR-07605 \citep{so14b}.
Although the onset/cease of the modulation seems to influence the pulsation period, its extent is too small to explain the anomalous period ratios of the Blazhko RRd stars, and the too long fundamental-mode periods of V44. 

Period changes of  the order of about one hundredth of a day have been detected  in some of the Blazhko RRd stars. A detailed analysis of their long-term behavior shows that the different components of the multiplets are the largest-amplitude signals at different epochs in these cases (Jurcsik et al. in preparation). However, none of the possible pairs of the components of the multiplets at the fundamental and overtone modes  yields a reasonable period ratio. The question, why and how the modulation influences the periods, hence the period ratios, remains to be solved. The detection of Blazhko modulation in RRd stars warns, that similarly to other Blazhko stars, the pulsation models fail to describe these stars properly.

The Blazhko statistics of the different samples of RR Lyrae stars shown in Sect 4 raise the question that why are  so few Blazhko stars among overtone variables? And why are most of the Blazhko RRc stars evolved,  over-luminous objects?

In the last decades, evidences have came to light that the star formation in GCs cannot be described as a single episode. The  chemical inhomogeneities and anomalies detected in many recent globular-cluster studies  \citep[for a review see ][]{g12} indicate a complex evolutionary history.  The observed variations in the abundances of the helium and the proton-capture process elements are supposed to be the results of self-enrichment processes connected to the multiple star-formation episodes during the early evolution of the GCs. Modeling the HB of the GCs has led to the conclusion that a surprisingly high  fraction ($20-100$\%) of the samples are chemically anomalous, helium-enriched second-generation objects \citep{ant}. The complex evolutionary and dynamical history of the GCs is the clue to understand the structure of the HB and the distribution of RR Lyrae stars, as well as  their properties within the instability strip (IS).

Modeling the color distribution of the HB stars and the period distribution of the RR Lyrae stars in M3, \cite{cda} argued that its blue HB consists mainly from He-enhanced, second-generation stars, and that most of the RRc stars belong to this population. According to their study, RRc stars with a period shorter than $P_{\mathrm{1O}}=0.335$~d ($P_{{\mathrm F}}=0.45$~d) or at $B-V<0.40$~mag are all helium enhanced.
Based on the $UV$ color-magnitude diagram of M3, the $Y$ and the mass of the HB stars are indeed different for stars with blue and red   colors \citep[see figure 10. in][]{d13}. Stars at  $\mathrm{m_{F336W}-m_{F555W}}>0.3-0.4$ have $Y=0.25$ and $M=0.63-0.70~M_\Sun$ values, while $Y=0.26$ and $M=0.55-0.63~M_\Sun$  fit the observations at hotter temperatures. 
In a recent study, \cite{jang} modeled the Oosterhoff properties of GCs assuming their different contents of first-, second-  and third-generation stars. They have concluded that the IS of an M3-like cluster is populated mostly by first-generation stars, but at the blue edge, third-generation objects with larger He content and smaller mass are also found. Contrary to these results, an upper limit of 0.01 for the He enhancement of the blue HB of M3 was given by \cite{cat}.

Supposing that there is no spread in the metallicity of the variables, their positions  on the HRD depend basically on their mass and HB evolutionary stage. 
The Blazhko RRd stars, close to the ZAHB magnitude,  are, most probably  either still close to their ZAHB position or they are evolving blueward. However, as V79 shows, this evolution is not always smooth and continuous. The similar Blazhko statistics of RRab and RRd stars points to that either both RRd and RRab stars have a property that makes them susceptible  for the Blazhko instability equally, or that the  RRd stars have evolved from RRab stars,   inheriting their modulations. 

 Presumably, the bright RRc stars with Blazhko statistics similar to the statistics of RRab/RRd stars  are already evolved objects. 
As the mass of the  second-generation stars is smaller than the mass of the first-generation objects, their evolution is slower than the evolution of the first-generation stars. Therefore, we suppose that the evolved RRc stars belong  to the first generation, too.

We thus conclude that,  the groups of RR Lyrae stars with Blazhko statistics around 50\% (RRab, RRd, and the over-luminous RRc sample) may belong to the first-generation of the cluster. Following the studies of \cite{cda,d13} and \cite{jang}, we propose that the normal brightness stars hotter than $B-V\approx0.35$~mag, i.e., the  RRc stars which do not show the Blazhko effect, are members of the second/third generation of the cluster. Based on the modest helium enhancement and mass difference of this subsample indicated by the $UV$ study, supposedly they are second-generation stars.

If this is indeed the case, then the appearance of the Blazhko modulation in M3 is connected to the evolutionary and chemical history of the stars, as the modulation occurs mostly (or exclusively) in first-generation objects.
There are other indications supporting that the Blazhko effect  occurs preferentially in first-generation stars, as well. 
Only second-generation stars are proposed to exist in M13 \citep{ant} and only first-generation ones in M53 \citep{ca}. In line with our idea, there is only one Blazhko star (showing a closely placed frequency) among the 9 variables of M13 \citep{kop} while 66\% and 37\% of the RRc and the RRab stars show the Blazhko effect in M53 \citep{af}, respectively. (The situation is, however, somewhat controversial, as  \cite{sand}  have found primordial He abundance in M13, and \cite{dk} determined a low Blazhko percentage in M53.)
The generation differences may also answer why the Blazhko effect is less frequent in overtone variables than in fundamental-mode stars \citep{mp} like in M4 \citep{m4}, in the SMC \citep{so10b} and in the  Galactic bulge \citep{so11}.
As the second/third-generation stars occupy bluer loci on the HB than first-generation stars \citep{dant,g12},  a larger fraction of RRc stars are suspected to  belong to the second-generation than of RRab stars. 

\acknowledgments
We thank G\'eza Kov\'acs and Aldo Valcarce for fruitful discussions about the results.
The authors are grateful to J. Kelemen, T. Kov\'acs, L. Kriskovics, E.
Kun, A. Mo\'or, A. P\'al, K. S\'arneczky, \'A. S\'odor, and J. Vink\'o,    
who obtained much of the data used in this analysis. The telescope and
camera control software package for the 90/60 Schmidt telescope was
developed by A. P\'al.
Support for G.H. is provided by the Ministry for the Economy, Development, and
Tourism's Programa Iniciativa Cient\'{i}fica Milenio
through grant IC\,210009, awarded to the Millennium Institute of Astrophysics (MAS);
by Proyecto Basal PFB-06/2007; and by Proyecto
FONDECYT Regular \#1141141.

\end{document}